\documentstyle[12pt,epsfig]{article}

  \newcommand{\ccaption}[2]{
    \begin{center}
    \parbox{0.85\textwidth}{
      \caption[#1]{\small{{#2}}}
      }
    \end{center}
    }
\catcode`\@=11

\def\nn{\nonumber}
\def\be{\begin{equation}}
\def\ee{\end{equation}}
\def\ba{\begin{eqnarray}}
\def\ea{\end{eqnarray}}

\def\Im{{\mathrm{Im}}}

\def\mev{\mbox{$\mathrm{MeV}$}}
\def\gev{\mbox{$\mathrm{GeV}$}}
\def\pt{\mbox{$p_T$}}

\def\chic{\mbox{$\chi_c$}}
\def\chij{\mbox{$\chi_J$}}

\def\hone{\mbox{$H_1$}}
\def\height{\mbox{$H_8$}}

\def\as{\mbox{$\alpha_s$}}
\def \oacube {\mbox{$ O(\alpha_s^3)$}}
\def \oatwo {\mbox{$ O(\alpha_s^2)$}}

\def \chiz {\mbox{$\chi_{0}$}}
\def \chio {\mbox{$\chi_{1}$}}
\def \chit {\mbox{$\chi_{2}$}}
\def \qq {\mbox{$Q \overline Q$}}

\def \chitolh {\mbox{$\Gamma(\chij \to LH)$}}

\def \chijqqg{\mbox{$\Gamma(\chi_J \to q \overline q g)$}}

\def \rprime {\mbox{$\vert R^{\prime}(0) \vert^2$}}
\def \ei {\mbox{$ \epsilon_{ \!\!\mbox{ \tiny{IR} } } $}}
\def \eu {\mbox{$ \epsilon_{ \!\!\mbox{ \tiny{UV} } } $}}
\def \s0 {\mbox{$\sigma_{0} $}}

\def \ep {\mbox{$\epsilon $}}
\def \cf {\mbox{$ C_F $}}
\def \ca {\mbox{$ C_A $}}
\def \caf {\mbox{$ C_F-\frac{1}{2}C_A $}}
\def \tf {\mbox{$ T_F $}}
\def \nf {\mbox{$n_{ \!\,\mbox{\tiny{Lf} } }  $}}
\def\fe{\mbox{$f(\ep)$}}
\def\der{\mbox{$\stackrel{\leftrightarrow}{\bf D}$}}
\def\nder{\mbox{$\stackrel{\leftrightarrow}{D}$}}

\begin{document}

\begin{titlepage}
\nopagebreak
{\flushright{
        \begin{minipage}{4cm}
        CERN-TH/96-84  \hfill 
      \\  hep-ph/9603439\hfill \\
        \end{minipage}        }

}
\vfill
\begin{center}
{\LARGE { \bf \sc  Colour-Octet \\ \vspace{.1cm}
                   NLO QCD Corrections\\ \vspace{.3cm} 
                   to Hadronic $\chi_J$ Decays}}
\vskip .8cm
{\bf Andrea PETRELLI}
\\
\vskip .1cm
{CERN, TH Division, Geneva, Switzerland and} \\
{Dipartimento di Fisica dell'Universit\`a and INFN, Pisa, Italy}
\end{center}
\vfill
\begin{abstract}
  In this paper we present a complete next-to-leading order QCD
  calculation of the \chij (\mbox{$^3P_J\; ;J=0,1,2 $}) hadronic
  decay width. We include  the NLO colour-octet contribution, as
  defined in the  Bodwin, Braaten and Lepage formalism. We
  extract an estimate of the colour-octet parameter \height\ from the
  charmonium decay data.
\end{abstract}
\vskip 1cm
CERN-TH/96-84  \hfill \\
March 1996 \hfill
\vfill
\end{titlepage}

{\bf 1.} Heavy quarkonium (HQ) systems are among  the most
interesting objects that nature gave us to explore perturbative
Quantum Chromodynamics (QCD).  The predictions of their production
cross sections and decay rates were among the most important tests of
the early-time QCD.  Nowadays there is a large renewed interest in the
physics of heavy quarkonium, due above all to the recent discovery of
the surprisingly big discrepancies between data and theory in the
high-\pt\ charmonium cross-section production at the Tevatron
\cite{CDFpsi}.  Bodwin, Braaten and Lepage (BBL) \cite{bbl} 
recently developed a new
formalism based on non-relativistic QCD (NRQCD) \cite{nrqcd} to
implement in a systematic way both the relativistic and the QCD
corrections to the naive Colour Singlet Model (CSM) (for a recent
review, see for example \cite{schuler}).  This framework has been
successfully applied to solve the charmonium anomaly at the Tevatron
\cite{tev1-cho}. Another crucial test of perturbative QCD in the heavy
quarkonium systems is given by the P-wave hadronic decay rates. In the
CSM the inclusive hadronic decay rate of P-wave HQ states shows a
singular infrared behaviour \cite{barbieri} which is a clear signal that such a  process
is sensitive to at least another non-perturbative parameter beyond the
usual wave function. In particular, the infrared problem of \chij\ 
($^3P_J$) decay arises from the \chijqqg\ subprocess. The amplitude
associated to this process diverges when the final gluon becomes soft.         This ambiguity spoils the traditional factorization
picture even at leading order in \as\ in the decay of the $\chi_1$
state into light hadrons (LH).  In the BBL theory there is the
solution of the \chij\ decay problem. In the CSM, the heavy quark pair
that participates in the hard annihilation process is in a colour
singlet state and has the same quantum numbers as the physical bound
state: the non-perturbative transition changes neither the colour nor
the spin-parity of the heavy-quark pair. In this picture the \chij\ 
decay occurs through the non-perturbative transition $ \chij\to Q
\overline Q[^3P_J^{(1)}]$ (the upper right label indicates the colour
state), which is parametrized by the derivative of the wave function
\rprime, followed by the annihilation of the $ Q \overline
Q[^3P_J^{(1)}]$ heavy quark pair.  Bodwin, Braaten and Lepage
suggested that the HQ wave function contains a non-negligible
component in which the heavy quark pair is in a $Q\overline
Q[^3S_1^{(8)}]$ state. This component leads to the HQ decay through
the process $Q\overline Q[^3S_1^{(8)}]\to q \overline q$. The
final state created by the colour-octet contribution is degenerate with
the colour-singlet one in the kinematical region of soft final gluon.
The colour-octet long-distance matrix element absorbs the infrared
sensitivity of the colour-singlet term yielding an IR-finite result
\cite{bblchi}.  In the NRQCD framework it is therefore possible to
give a theoretical prediction of $\chi_{J}$ hadron decays avoiding
infrared inconsistencies.

Hadronic \chij\ annihilation then gives a very important
phenomenological test of the role of the colour-octet mechanism, and of
the BBL theory in general.  In ref.~\cite{bblchi} Bodwin, Braaten and
Lepage performed a phenomenological analysis of $\chic_J$ decays using
the LO results for both the colour-singlet and the colour-octet
contributions. They justified the neglect of the known NLO colour-singlet     
corrections with the observation that NLO accuracy would
require inclusion of the yet unknown NLO colour-octet coefficients. In
a following paper \cite{chi} it was argued that it is justified to
include in the analysis the available QCD NLO colour-singlet terms,
because the octet
contribution does not depend on $J$ even at NLO. Therefore the 
higher-order colour-octet coefficient can be simply reabsorbed in a redefinition
of the LO octet wave function without changing the values of \as\ and
\rprime\ extracted from the fit to the experimental data.

In this work we perform 
the NLO calculation of the colour-octet coefficient completing the
picture of the \chij\ hadron decay at order \mbox{$ O(\as^3 v^5)$} in
the sense of the BBL double expansion. The knowledge of the NLO octet
correction allows us to extract the value of the  parameter \height\ 
using the results of ref~\cite{chi}.\newline An analogous calculation relative to the newly
discovered \mbox{$^1P_1$} state \cite{E760-h} has recently appeared \cite{cina}.\newline

{\bf 2.} The \chij\ state is represented as a Fock-space vector superposition of heavy quark pair states of different spin, angular momentum and colour,  possibly accompanied by gluons \cite{bbl}: 
\ba
\label{chiexp}
\vert\chi_{J}\rangle =O(1)\vert\qq [^3 P_J^{(1)}]\rangle
+O(v)\vert\qq [^3 S_1^{(8)}] g\rangle + \cdots ,
\ea
where $v$ is the relative velocity between the bound quarks.  
The first term of eq. (\ref{chiexp}) represents the conventional colour-singlet configuration and the second one corresponds to a colour-octet heavy quark pair in the \mbox{$Q\overline Q[^3S_1^{(8)}]$} state accompanied by a gluon.  
The small relative bound quark velocity $v$  splits the physics of heavy quarkonium into two well separated energy scales, allowing a formal factorization of the physical observables into perturbative short-distance kernels describing annihilation of the heavy quark pair and soft non-perturbative coefficients. In the BBL factorization framework, the low energy heavy quarkonium physics is described by the NRQCD Lagrangian which has a physical ultraviolet cutoff $\Lambda $~. 
The short-distance annihilation effects are implemented including  4-fermion interactions in the Lagrangian:
\ba
\label{4f}
\delta {\cal L}_{4-fermions} = \Sigma_n
\frac{f_{n}(\Lambda)}{m^{\delta_n - 4}} {\cal O}_{n}(\Lambda),
\ea
where $m$ is the mass of the heavy quark.
Both the  NRQCD operators ${\cal O}_{n}$ and the short-distance coefficients $f_n$ depend on $\Lambda$, but their product does not.
The operators  ${\cal O}_{n}$ have well-defined scaling rules with velocity $v$ and  the coefficients $f_n$ have a QCD perturbative definition. Equation (\ref{4f}) can be actually read as a double \as\ and $v$ expansion. 
For our study, the relevant operators ${\cal O}_n$ are: 
\clearpage
\[
\]
\ba
&&{\cal O}_1(^3P_0)=\frac{1}{3}\psi^{\dagger}\left(-\frac{i}{2}\der\cdot{\bf \sigma}\right )\phi\;\phi^\dagger\left(-\frac{i}{2}\der\cdot{\bf \sigma}\right)\psi\\
&&{\cal O}_1(^3P_1)=\frac{1}{2}\psi^{\dagger}\left(-\frac{i}{2}\der\times{\bf \sigma} \right)\phi\cdot\phi^\dagger\left(-\frac{i}{2}\der\times{\bf \sigma}\right)\psi\\
&&{\cal O}_1(^3P_2)=\psi^\dagger \left(-\frac{i}{2}\nder\, ^{(i}\sigma^{j)}\right)\phi\;\phi^\dagger \left(-\frac{i}{2}\nder\,^{(i}\sigma^{j)}\right)\psi\\
&&{\cal O}_8(^3S_1) = \psi^\dagger {\bf \sigma}T^a\phi\cdot\phi^\dagger{\bf\sigma}T^a\psi
\ea 
\[
\] 
The \chij\ hadronic decay width can be written as: 
\[
\]
\ba
\Gamma(\chi_J\to LH)=2\; \Im f_1(^3P_J)\; \frac{ \langle\chi_J\vert{\cal
O}_1(^3P_J)\vert\chi_J\rangle}{m^4} + 2\; \Im f_8(^3S_1)\;\frac{
\langle\chi_J\vert{\cal O}_8(^3S_1)\vert\chi_J\rangle}{m^2} 
\ea
\[
\]
The short-distance coefficients can be extracted by matching NRQCD and full QCD amplitudes \cite{bbl}. The NRQCD matrix elements can be determined phenomenologically or calculated on the lattice. 
\[
\]
Defining:
\ba
\label{h1h8}
\hone = \frac{ \langle\chij \vert{\cal O}_1(^3P_J)\vert\chij \rangle}{m^4}&& 
\;\;\;\;\height(\Lambda) = \frac{\langle\chij \vert{\cal O}_8(^3S_1;\Lambda)\vert\chij \rangle}{m^2}. 
\ea
\[
\]
we can rewrite the \chij\ width as follows
\[
\]
\ba
\chitolh = \hat\Gamma_1(^3P_J^{(1)}\to LH)\hone+\hat\Gamma_8(^3S_1^{(8)}\to LH)\height = \hat\Gamma_1(J) \hone +\hat\Gamma_8 \height
\ea
\[
\]
Velocity and mass scaling of the matrix elements of the relevant NRQCD operators are \mbox{$\hone\sim m v^5$}, 
\mbox{$\height\sim m v^5$}.  The QCD leading-order colour-singlet short-distance coefficients are of \oatwo\ for \chiz\ and \chit\, and of \oacube\ for \chio\ states. On the other hand, the QCD lowest order colour-octet short-distance process  
\mbox{$\qq  [^3S_1^{(8)}]  \to q \overline q $} is of  \oatwo\, while the process \mbox{$\qq [^3S_1^{(8)}]\to  g g $} with both gluons on the mass shell is forbidden. 
Therefore a consistent perturbative picture of the \chio\ decay at order ${\as}^3$$v^5$ requires the calculation of the QCD NLO colour-octet contribution. 
For ease of reference, we collect here the expression for the \chij\ decay widths including the NLO colour-singlet terms \cite{barbieri} and the LO colour-octet terms \cite{bblchi}
\[
\]
\be
\begin{array}{lll}
\label{singlet}
\Gamma(\chiz \to LH) = \frac{4}{3}  \pi \as^2 \hone\Large[1\; +\! &\frac{\as}{\pi} C_0 \Large] &+\;\; \nf \frac{\pi}{3}\as^2\left[\frac{16}{27}\frac{\as}{\pi}\hone\log\frac{m}{\cal E} + \height \right] \\
& & \\
& & \\
\Gamma(\chit \to LH) = \frac{4}{3}  \pi \as^2 \hone\Large[1\; +\!& \frac{\as}{\pi} C_2 \Large]\frac{4}{15} &+\;\; \nf \frac{\pi}{3}\as^2\left[\frac{16}{27}\frac{\as}{\pi}\hone\log\frac{m}{\cal E} + \height \right] \\
& & \\
& & \\
\Gamma( \chi_1 \to LH) = \frac{4}{3}\pi\as^2\hone\Large[ &\frac{\as}{\pi} C_1 \Large] &+\;\; \nf\frac{\pi}{3}\as^2\left[\frac{16}{27}\frac{\as}{\pi}\hone\log\frac{m}{\cal E} + \height \right]  
\end{array}
\ee
\[
\]
where
\[
\]
\be
\begin{array}{l}
C_0 = \left(\frac{454}{81} -\frac{1}{144}\pi^2 -\frac{11}{3}\log
2\right)\ca +\left( -\frac{7}{3} + \frac{\pi^2}{4}\right)\cf +\nf\left( -\frac{16}{27}+ \frac{2}{3}\log 2 \right) \\
\\
\\
 C_2 = \left( \frac{2239}{216} -\frac{337}{384}\pi^2 -2 \log 2 \right)\ca -4\cf+\nf\left( -\frac{11}{18}+\frac{2}{3}\log 2 \right)\\
 \\
\\
C_1=\left(\frac{587}{54}-\frac{317}{288}\pi^2\right)+\nf\frac{28}{81}
\end{array}
\ee
\[
\]
\nf\ is the number of  light quarks: \nf\ =\ 3 for charmonium and \nf\ =\ 4 for bottomonium states and \as\ = \mbox{$ \as (m) $}. \hone\ is related to the derivative of the wave function through the relation:\
\ba
\hone = \frac{9}{2\pi} \frac{\rprime}{m^4}\left[1+O(v^2)\right]
\ea    
\[
\]
We denote by ${\cal E}$ a momentum  scale that regularizes the soft divergence associated to the \mbox{$\chij\to q \overline q g$} process. $\cal E$ was usually related to the binding energy of quarkonium. 
Notice that neither the colour-octet term nor the coefficient of the divergent logarithm depends on the quarkonium spin $J$; this fact makes  a universal renormalization of the parameter \height\ possible. In fact, considering  only the universal piece (\mbox{$U_{\Gamma}$}) of \chij\ widths we get: 
\[
\]
\ba 
U_{\Gamma}&=&\nf \frac{\pi}{3}\as^2 \left[\frac{16}{27}\frac{\as}{\pi}\hone\left(\log\frac{m}{\Lambda}+\log\frac{\Lambda}{\cal E}\right) + \height^{(b)} \right] \\  
&=&\nf \frac{\pi}{3}\as^2 \left[ \height (\Lambda) +\frac{16}{27}\frac{\as}{\pi}\hone\log\frac{m}{\Lambda}\right] = \nf \frac{\pi}{3}\as^2 \height (m).  
\ea
\[
\]
The \mbox{$\Lambda$}-dependence of the colour-singlet coefficient is consistent with that specified by the RGE for \height\ \cite{bbl}.  
\[
\]
\[
\]
{\bf 3.}
We perform the calculation of the full NLO QCD colour-octet contribution to the \chij\ decay widths term using the dimensional regularization scheme to regularize UV, IR and collinear divergences. We work in D$=$ 4 $-$ 2$\epsilon$ dimensions. 
If we define:
\ba
\hat\Gamma_8^{(0)} = \pi\as^2\frac{1-\ep}{3-2\ep}\left(\frac{4\pi\mu^2}{M^2}\right)^{\ep}\frac{\Gamma (1-\ep)}{\Gamma(2-2 \ep)}
\ea
where $M$ $\equiv$ 2 $m$, 
then the D-dim Born colour-octet short-distance coefficient assumes the form:
\ba
\hat\Gamma_8^{(Born)}=\nf\hat\Gamma_8^{(0)}
\ea
The NLO correction to \mbox{$\hat\Gamma_8$} consists of real and virtual emission of gluons.\newline
\[
\]
{\em Real emission\/}\newline

The real correction to the short-distance colour-octet \chij\ 
annihilation term is represented by the two processes \mbox{$ \qq
  [^3S_1^{(8)}]\to g g g$} and \mbox{$ \qq [^3S_1^{(8)}]\to q
  \overline q g$} .  The calculation of the D=4 \mbox{$ \qq
  [^3S_1^{(8)}]\to g g g$} amplitude can be obtained via crossing from
the results of ref \cite{cho}. This amplitude is completely IR and
collinear finite because the two-gluon leading order amplitude
vanishes. The calculation of the three-gluon real contribution is
therefore straightforward.  We obtain \footnote{The expression
  reported here assumes implicitly $N_c=3$, since the explicit $N_c$
  dependence of the matrix elements is not reported in the result of
  ref.~\cite{cho}}:
\[
\]
\ba
\hat\Gamma_8^{(ggg)} = \hat\Gamma_8^{(0)}\frac{\as}{\pi}\; 5\;\left(-\frac{73}{4}+\frac{67}{36}\pi^2\right)
\label{ggg}
\ea
\[
\]
On the contrary  the process \mbox{$ \qq [^3S_1^{(8)}]\to q \overline q g$}  
shows IR and collinear poles that we expect will cancel when adding the virtual correction. 
The D-dimension \mbox{$\qq [^3S_1^{(8)}]\to q\overline q g$} amplitude that we obtain is in agreement with ref~\cite{cho} in the D$=$4 limit. It leads to the following width contribution 
\[
\]
\ba
d\hat\Gamma_8^{(q\overline q g)}=\nf\frac{64 \pi^3\as^3}{M^2(3 - 2 \ep)}\left[t^2+u^2+2M^2s-\epsilon (t+u)^2\right]\left[\frac{\cf}{t u} - \frac{\ca}{(s-M^2)^2}\right] d(PS)[q\overline q g]
\ea
\[
\]
where \mbox{$s=(q+\overline q)^2$}, \mbox{$t=(q+g)^2$}, \mbox{$u=(\overline q +g)^2$}. 
Performing the Mandelstam variable substitution \mbox{$ s=M^2(1-x)$}, \mbox{$ t=M^2 x y$}, \mbox{$ u=M^2 x(1-y)$}, the phase space assumes the following form: 
\[
\]
\ba
d(PS)[q\overline q g] = \frac{M^2}{128 \pi^3}\left(\frac{4\pi\mu^2}{M^2}\right)^{2\ep} \frac{1}{\Gamma (1-\ep)} x\left[x^2 (1-x) y(1-y)\right]^{-\ep}dx dy.
\label{ps}
\ea
\[
\]
Integrating over the phase space we get
\[
\]
\ba
\hat\Gamma_8^{(q\overline q g)} =\nf\hat\Gamma_8^{(0)}\frac{\as}{\pi}\fe    \left[\cf \left(\frac{1}{\ep^2}+\frac{3}{2\ep}\right) +\ca\frac{1}{2\ep} +\cf\left(\frac{19}{4}-\frac{2}{3}\pi^2\right) +\frac{11}{6}\ca\right]
\label{qqg}
\ea
\[
\]
where
\ba
\fe = \left(\frac{4\pi\mu^2}{M^2}\right)^{\ep}\Gamma(1+\ep)\;\; . 
\ea
\[
\]
\[
\]
{\em Virtual emission\/}\newline
The diagrams contributing to NLO virtual emission are shown in fig. (1). In table 1 we list the contribution \mbox{${\cal D}_{k}$} of each diagram with the relative colour factors. The virtual colour-octet width can be written as
\ba
\hat\Gamma_{8,QCD}^{(Virtual)} = \nf\hat\Gamma_8^{(0)}\fe\frac{\as}{\pi} {\sum}_k({\cal D}_k f_k)
\ea
Summing all the virtual diagrams, we find 
\[
\]
\ba
{\hat \Gamma}_{8,QCD}^{(Virtual)} &=& \nf{\hat \Gamma}_{8}^{(0)}\left\{ 1 + \frac{\as}{\pi}\fe\left[ 2 b_0 \frac{1}{\eu} - \cf \left(\frac{1}{\ei^{\!\!\!2}}+\frac{3}{2\ei}\right) - \ca\frac{1}{2\ei}\right.\right.\nn\\
&+&\left.\left.\frac{\pi^2}{v}\left(\cf-\frac{1}{2}\ca\right) + A \right] \right\} 
\ea
\[
\]
where
\[
\]
\ba
&&b_0 = \frac{1}{2}\left(\frac{11}{6}\ca- \frac{2}{3}\nf\tf\right) \\
&&A= \cf\left(-8+\frac{2}{3}\pi^2\right) + \ca\left( \frac{50}{9} +\frac{2}{3}\log 2 -\frac{\pi^2}{4}\right) -\frac{10}{9}\nf\tf\\
&&v\equiv \left[ 1- \frac{4 m^2}{(p_Q + p_{\overline Q})^2}\right]^{-\frac{1}{2}}.
\ea
\[
\]
The Coulomb singularity disappears by performing the matching between the NLO full QCD and NRQCD amplitudes, yielding a finite result in the $v\to\,0$ limit.  
Summing the real and the virtual emission corrections, we obtain the \as\ NLO colour-octet decay width for \chij\ states; we give the result for \cf\ =4/3,      \ca\ $=$ 3 and \tf\ $=$1/2: 
\[
\]
\ba
{\hat \Gamma}_{8}^{(NLO)} &=& \hat\Gamma_8^{(Born)}\left[ 1 + \frac{\as^{\overline{MS}}(\mu )}{\pi}\left(\; \frac{107}{6}-\frac{3}{4}\pi^2 + 2\log 2\; +\right.\right.\nn\\ &-&\left.\left. \frac{5}{9}\nf + 4 b_0\log\frac{\mu}{2\,m}\; \right) \right] + \hat\Gamma_8^{(0)} \frac{\as^{\overline{MS}}(\mu )}{\pi}\;5\;\left(-\frac{73}{4}+\frac{67}{36}\pi^2\right).
\ea
\[
\]
Choosing the renormalization scale \mbox{$\mu = m$} ( the same of the colour-singlet terms in eqs.~(\ref{singlet})) we obtain the following expression of the \as\ NLO imaginary part of $f_8(^3S_1)$:
\[
\]
\ba
\Im\;f_8(^3S_1) &=&\frac{\pi}{6}\left({{\as}^{\overline {MS}}}(m)\right)^2 \left[ \nf + \nf\frac{\as^{\overline {MS}}(m)}{\pi}\left(\;\frac{107}{6}-\frac{3}{4}\pi^2 - 9 \log 2\; +\right.\right.\nn\\&-&\left.\left. \frac{5}{9}\nf +\frac{2}{3}\nf\log 2 \right) + \frac{\as^{\overline {MS}}(m)}{\pi}\,5\left(-\frac{73}{4}+\frac{67}{36}\pi^2\right)\right]
\ea  
\[
\]
Always keeping $\mu=m$, we report below the numerical colour-octet corrections for charmonium and bottomonium:
\[
\]
\ba
\frac{ {\hat \Gamma}_{8}^{(NLO)}}{\hat\Gamma_8^{(Born)}}= 1+3.9\;\frac{{\as}^{\overline {MS}}(m_c)}{\pi}   & &[{\mathrm charm}]\\
\frac{ {\hat \Gamma}_{8}^{(NLO)} }{\hat\Gamma_8^{(Born)}}= 1+3.8\;\frac{{\as}^{\overline {MS}}(m_b)}{\pi}  & &[{\mathrm bottom}] 
\ea
\[
\]
Using the results of ref. \cite{chi} it is straightforward to obtain
the best fit of the parameter \height\ for charmonium \chic\ including
the NLO QCD effects, and we get the value \mbox{$\height^{(c)}(m_c)$}
= 3.1 $\pm$ 0.5 \mev. Taking \mbox{$m_c=1.5\;\gev$} we obtain
\mbox{$\langle\chic_J \vert{\cal O}_8(^3S_1;m_c)\vert\chic_J \rangle
  $} $=$ \mbox{$ (6.8\;\pm\;1.1)\times 10^{-3}\;\;\gev^3 $}. For
completeness we recall that the fits of the other parameters obtained
in \cite{chi} are \mbox{$\hone = 13.7\,\pm\,2.3\;\mev$} and
\mbox{$\as^{\overline{MS}}(m_c)\, = 0.286\,\pm\,0.031$}. As discussed
in the introduction these results are not affected by the inclusion of
the NLO colour-octet corrections. Using the NRQCD scaling rules, we
can obtain an estimate of the bottom octet matrix element
$\height^{(b)} (m_b)$ $\simeq$ $0.66\;\mev$.
  
We now want to analyse the renormalization scale dependence of the NLO
colour-octet decay widths compared with the leading-order ones. The
results are shown in figs. 2 and 3 for charmonium and bottomonium
states, respectively. The normalization of the bottomonium width is
achieved by using the estimate of the colour-octet parameter
$\height^{(b)} (m_b)$ obtained above through the NRQCD scaling rules.
For the running of two-loop \as\ we use the input $\Lambda^{\overline
  {MS}}_{n_f=5}\;=\; 160\;\mev\;$ extracted from the fitted value of
$\as^{\overline{MS}}(m_c)$. The pictures show that the inclusion of
NLO corrections significantly reduces the scale dependence of the
processes.
 
To conclude we notice that the calculation presented here can be used
to compute the strong NLO \mbox{$q \overline q \to Q\overline Q
  [^3S_1^{(8)}]$} contribution to the total \mbox{$\psi $} hadronic
production cross section and the NLO colour-octet fragmentation
function of the gluon into $ \psi $. Work on these issues is in
progress.

\[
\]
{\renewcommand{\arraystretch}{1.8}
\begin{table}
\begin{center} 
\begin{tabular}{lll} \hline\hline

Diag. &${\cal D}_k$ &$f_k$ \\ \hline\hline

 a& $\left[ -\frac{1}{2 \eu} + \frac{1}{2 \ei}\right] $&
 $\cf$\\ \hline
 b& $\left[ \frac{1}{2
 \eu}-\frac{1}{{\ei}^{\!\!\!2}}-\frac{2}{\ei}-4+\frac{2}{3}{\pi}^2\right]$ &\caf \\ \hline
 c&  $\left[ \frac{3}{2 \eu}-\frac{2}{\ei}-1\right]$ &$\frac{1}{2}$\ca \\ \hline
 d & $ \left [-\frac{1}{2 \eu}-\frac{1}{\ei}-2-3\log
 2\right]$ &\cf \\ \hline
 e & $ \left [\frac{\pi^2}{v}+\frac{1}{2
\eu}+\frac{1}{\ei}- 2 + 3 \log 2\right] $ &\caf\\ \hline 
 f & $\left[ \frac{3}{2 \eu}
 +\frac{8}{3}+\frac{13}{3}\log 2\right]$&$\frac{1}{2}$\ca \\  \hline
 g & $ \left [\frac{5}{6}\frac{1}{\eu} +
\frac{31}{18}\right] $ &\ca\\  \hline
 h & $ \left[-\frac{2}{3 \eu} -\frac{10}{9}\right] \nf $&\tf\\  \hline
 i & $\left[\frac{1}{{\ei}^{\!\!\!2}} 
 - \frac{{\pi}^2}{6}\right] $ & $2\cf - \ca$\\ \hline
 j & $ \left[- \frac{1}{ {\ei}^{\!\!\!2} } + \frac{{\pi}^2}{6}\right] $ &$ 2\cf - \frac{1}{2}\ca$\\ \hline\hline
\end{tabular}
\caption{Partial virtual QCD corrections to the process $\qq  [^3S_1^{(8)}]  \to q \overline q $ }
\end{center}
\end{table}
\clearpage
\begin{center}
{\bf Acknowledgements }
\end{center}
It is a pleasure for me to thank Michelangelo Mangano for his
generosity and several enlightening suggestions. I wish to thank also
Matteo Cacciari, Vitaliano Ciulli for useful conversations and G.
Veneziano for his hospitality at the CERN Theory Division.

\begin{figure}[t]
\begin{center}
\psfig{figure=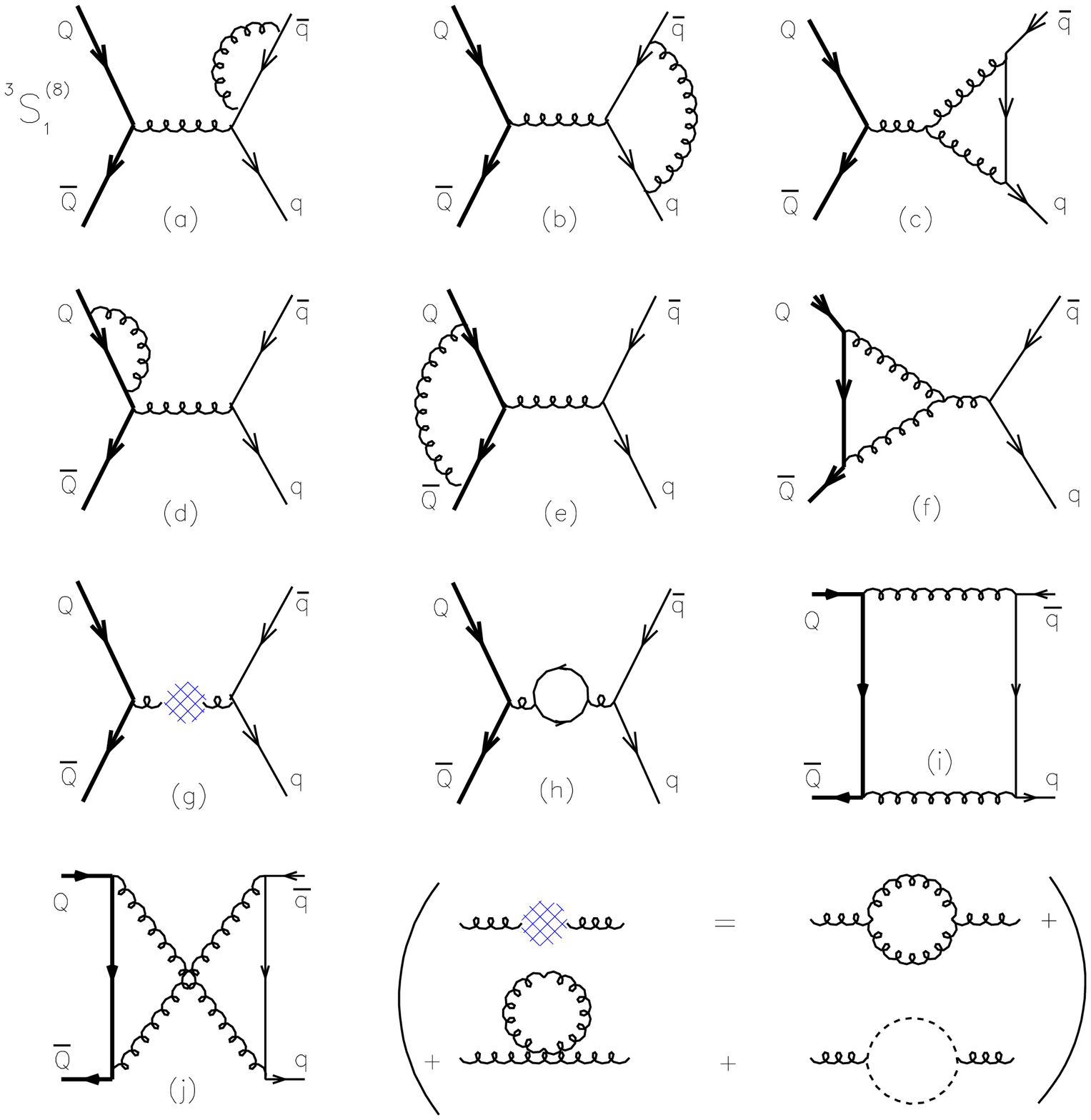,width=15cm,angle=0}
\ccaption{}{\small Virtual Feynman diagrams contributing to the process $\qq  [^3S_1^{(8)}]  \to q \overline q $ }
\end{center}
\end{figure}

\begin{figure}
\begin{center}
\psfig{figure=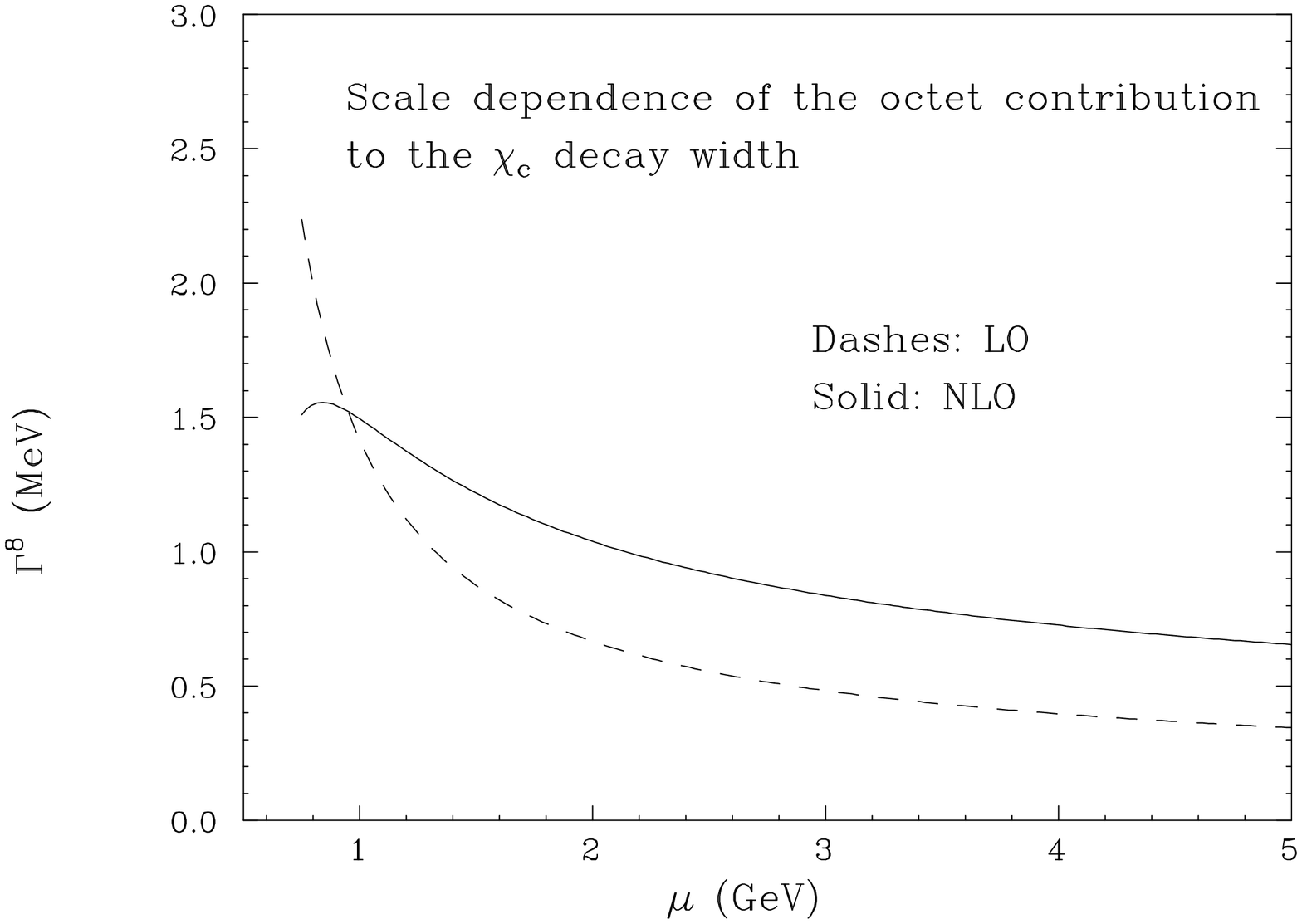,width=8cm,angle=90}
\ccaption{}{\small Renormalization scale dependence of the colour-octet contribution to $\chi_{c}$ hadronic decay width $\Gamma_8~=~\hat\Gamma_8\height$ }
\end{center}
\end{figure}
\begin{figure}
\begin{center}
\psfig{figure=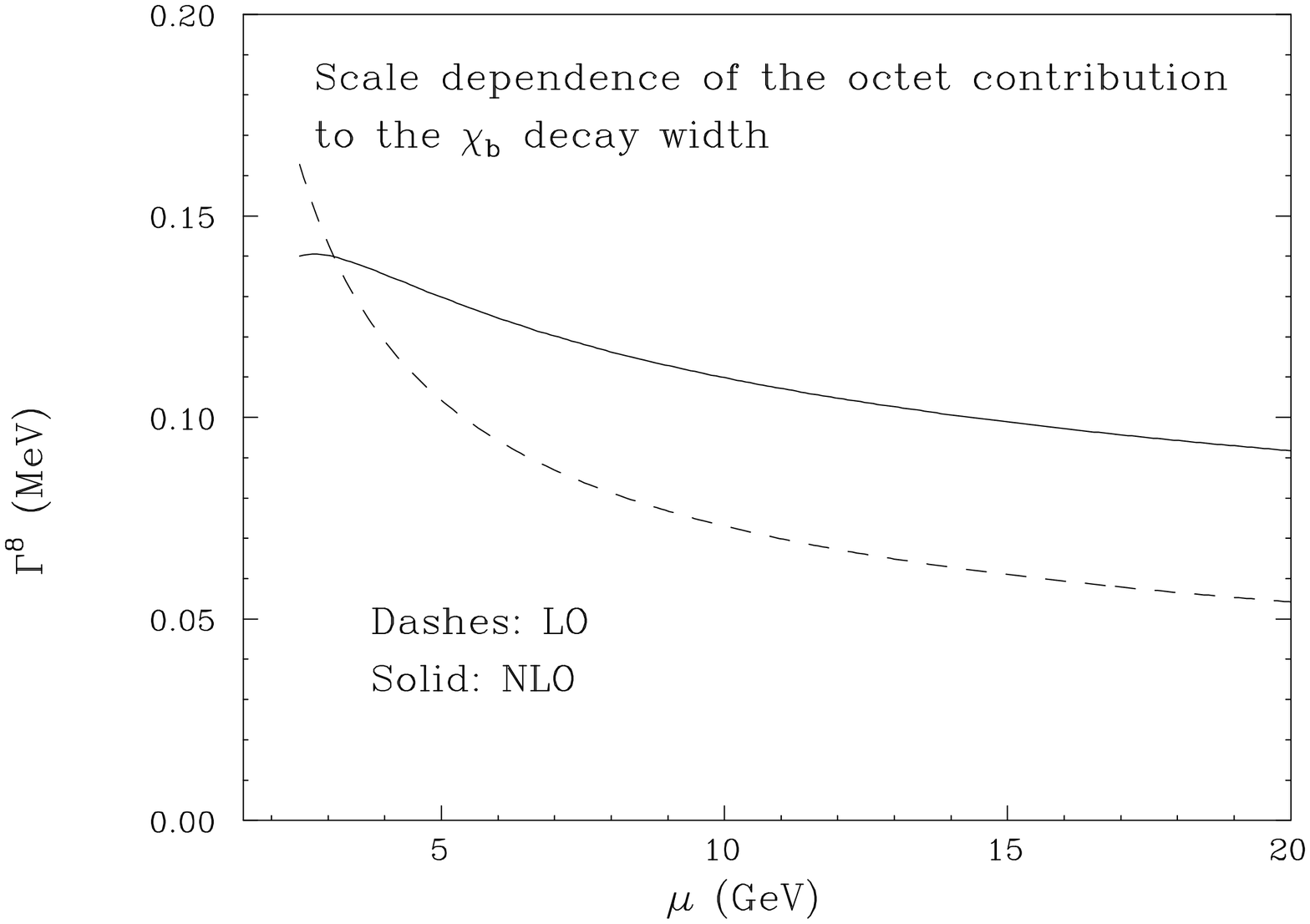,width=8cm,angle=90}
\ccaption{}{\small Same as fig. 2 but for $\chi_b$ }
\end{center}
\end{figure}

\end{document}